\documentclass{PoS}

\usepackage{url}

\title{Production of $^{92}$Nb, $^{92}$Mo, and $^{146}$Sm in the $\gamma$-process in SNIa}

\ShortTitle{Production of $^{92}$Nb, $^{92}$Mo, and $^{146}$Sm in SNIa}

\author{\speaker{T. Rauscher}$^{abc}$, C. Travaglio $^d$, R. Gallino $^e$, N. Nishimura $^{cf}$, R. Hirschi $^{cfg}$\\
\llap{$^a$} Centre for Astrophysics Research, University of Hertfordshire\\
Hatfield AL10 9AB, United Kingdom\\
\llap{$^b$} Department of Physics, University of Basel\\
4056 Basel, Switzerland\\
\llap{$^c$} UK Network for Bridging Disciplines of Galactic Chemical Evolution (BRIDGCE)\\
\url{http://www.astro.keele.ac.uk/bridgce}, United Kingdom\\
\llap{$^d$}INAF-Astrophysical Observatory Turin\\
10025 Pino Torinese (Turin), Italy\\
\llap{$^e$} Dipartimento di Fisica, Universit\'a di Torino\\
10125 Turin, Italy\\
\llap{$^f$} Astrophysics group, Keele University\\
Lennard-Jones Labs, Keele ST5 5BG, United Kingdom\\
\llap{$^g$} Institute for the Physics and Mathematics of the Universe (WPI), University of Tokyo\\
5-1-5 Kashiwanoha, 277-8583 Kashiwa, Japan
        }

\abstract{The knowledge of the production of extinct radioactivities like $^{92}$Nb and $^{146}$Sm by photodisintegration processes in ccSN and SNIa models is essential for interpreting abundances in meteoritic material and for Galactic Chemical Evolution (GCE). The $^{92}$Mo/$^{92}$Nb and $^{146}$Sm/$^{144}$Sm ratios provide constraints for GCE and production sites. We present results for SNIa with emphasis on nuclear uncertainties.}

\FullConference{XIII Nuclei in the Cosmos\\
                 7-11 July, 2014\\
                 Debrecen, Hungary}

\begin{document}

\section{Introduction}
The existence of now extinct radioactivities such as $^{92}$Nb and $^{144}$Sm in the early solar system can be inferred from meteoritic measurements \cite{dau11,dav13}. The obtained ratios $^{92}$Mo/$^{92}$Nb and $^{146}$Sm/$^{144}$Sm provide constraints for GCE and production sites but are still prone to nuclear uncertainties as only few of the important reactions are experimentally determined. We present calculations for a single-degenerate Chandrasekhar-mass SNIa model \cite{DDT}, in which material is accreted from a companion AGB star and enriched in s-process seeds before the explosion of the White Dwarf. The temperature and density history of a mass element is followed through the explosion by using tracers. We scrutinize specific tracers with maximal production of $^{92}$Mo, $^{92}$Nb, and $^{146}$Sm to discuss the nuclear uncertainties. The full details of calculation and tracer choice are described in \cite{trav14}.

\section{Production Paths and Uncertainties}
\label{sec:paths}

\begin{figure}
\includegraphics[width=0.5\columnwidth]{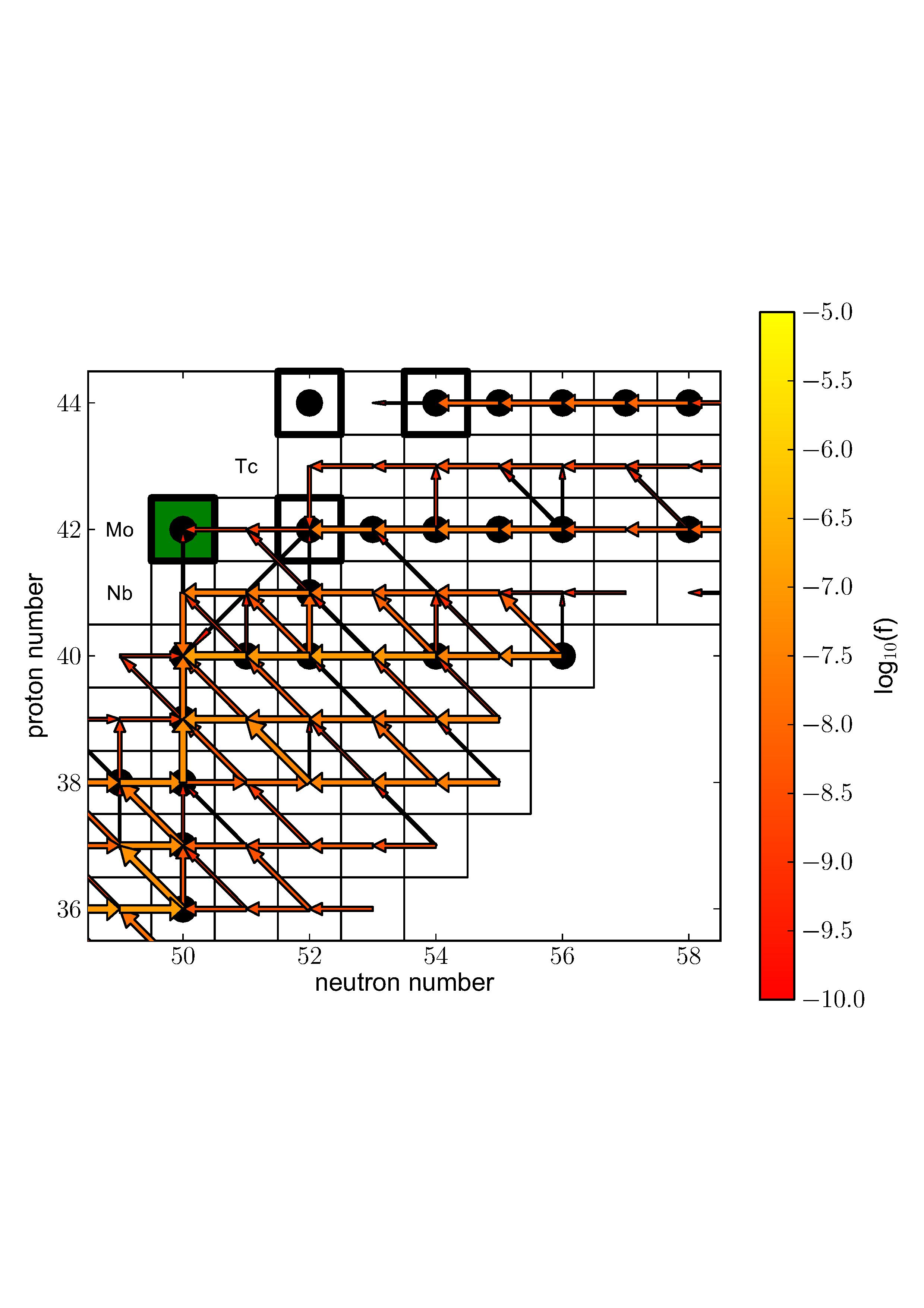}
\includegraphics[width=0.5\columnwidth]{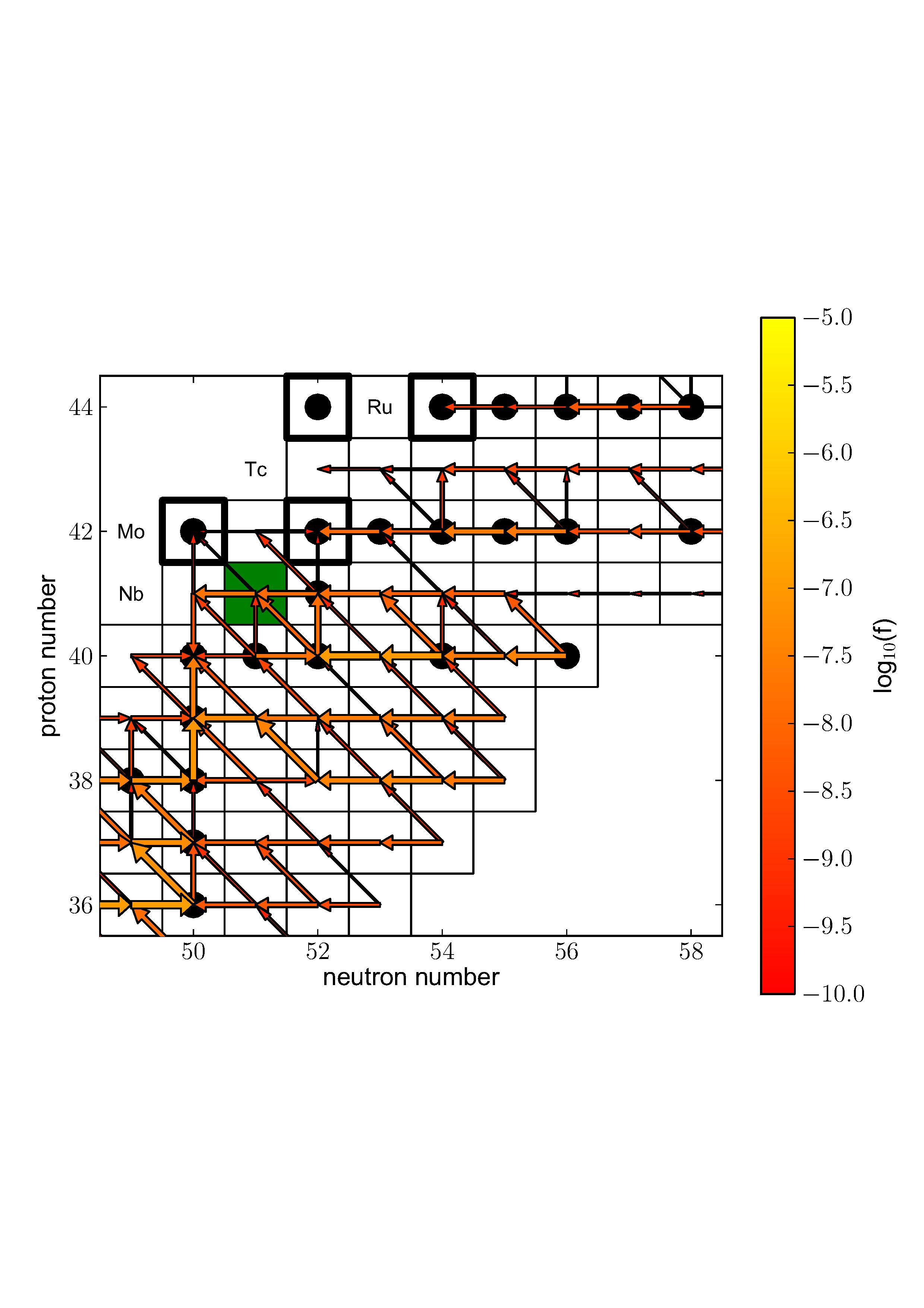}
\caption{\label{fig:flowMoNb}Reaction flow in the two SNIa trajectories leading to the maximal $^{92}$Mo (left) and $^{92}$Nb (right) production, respectively; size and color of the arrows relate to the magnitude of the time-integrated flux on a logarithmic scale.}
\end{figure}

\begin{table}
\begin{center}
\scriptsize
\begin{tabular}{ccc}
Reactions & Rate set MIN & Rate set MAX \\
\hline
$^{91}$Zr(p,$\gamma$)$^{92}$Nb & $\downarrow$ & $\uparrow$ \\
$^{92}$Zr(p,$\gamma$)$^{93}$Nb & $\downarrow$ & $\uparrow$ \\
$^{92}$Zr(p,n)$^{92}$Nb & $\downarrow$ & $\uparrow$ \\
$^{91}$Nb(n,$\gamma$)$^{92}$Nb & $\uparrow$ & $\downarrow$  \\
$^{92}$Nb(n,$\gamma$)$^{93}$Nb & $\downarrow$ & $\uparrow$\\
&& \\
$^{91}$Nb(p,$\gamma$)$^{92}$Mo & $\uparrow$ & $\downarrow$  \\
$^{93}$Nb(p,n)$^{93}$Mo & $\uparrow$ & $\downarrow$  \\
$^{93}$Mo(n,$\gamma$)$^{94}$Mo & $\uparrow$ & $\downarrow$ \\
\hline
GCE  & 1.66$\times 10^{-5}$ & 3.12$\times$10$^{-5}$
\end{tabular}
\caption{\label{tab:nbmoratio}Reactions affecting the $^{92}$Nb/$^{92}$Mo ratio and their
variation to explore the nuclear uncertainties; rate set MIN yields the minimal ratio, set MAX the maximal ratio. The arrows indicate whether a rate has been multiplied by a factor of two
(arrow up) or divided by the same factor (arrow down). The modifications always apply to the rate and its reverse rate. The resulting value for the $^{92}$Nb/$^{92}$Mo abundance ratio after performing a GCE calculation is also shown.}
\end{center}
\end{table}

\begin{figure}
\includegraphics[width=0.5\columnwidth]{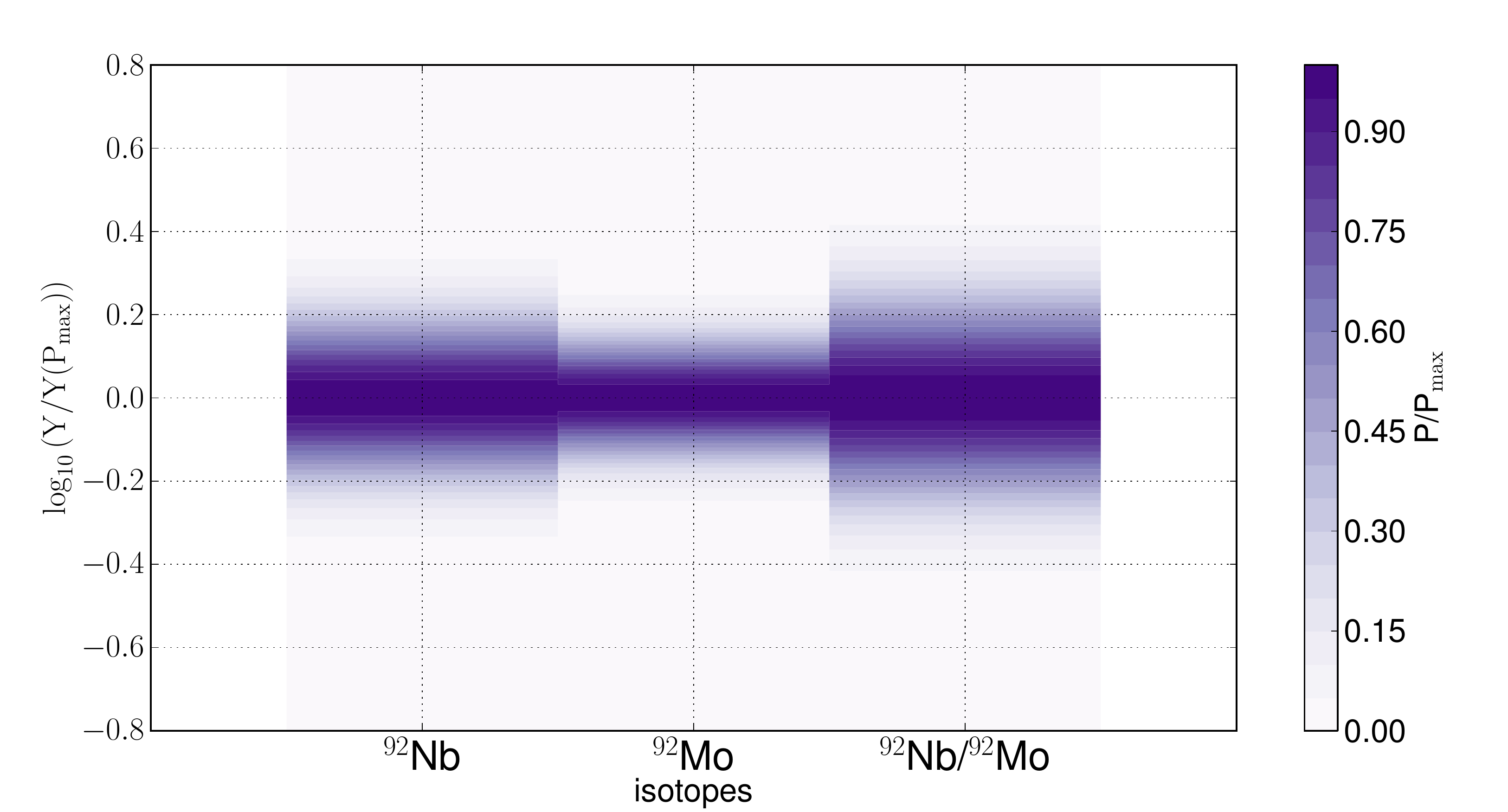}
\includegraphics[width=0.5\columnwidth]{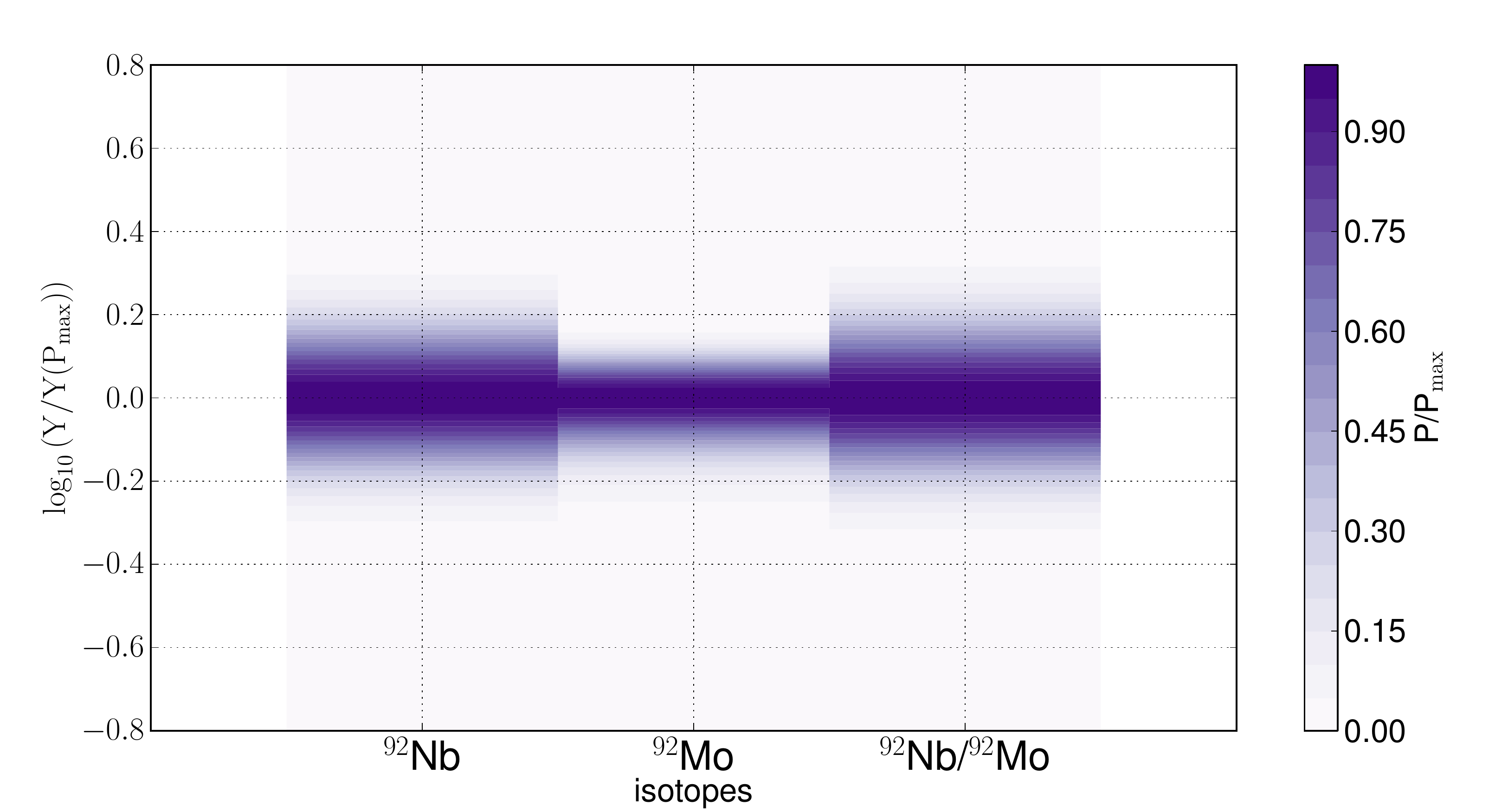}
\caption{\label{fig:dens}Probability density distributions (PDDs) for $^{92}$Nb (left), $^{92}$Mo (center) and their ratio (right) obtained from Monte Carlo variations of all reactions (left panel) and only the reactions given in Table \protect\ref{tab:nbmoratio} (right panel). For the MC variations, different uncertainty factors were used depending on the reaction and whether it is predicted or measured.}
\end{figure}

The production of the radiogenic $^{92}$Nb is governed by the destruction of $^{93}$Nb and $^{92}$Zr seeds, as can be seen from the flows in Fig.\ \ref{fig:flowMoNb}. It also gets some indirect contributions from $^{91,94,96}$Zr via $^{92}$Zr but none from $^{90}$Zr. The nuclide $^{92}$Nb is mainly destroyed by the reaction $^{92}$Nb($\gamma$,n)$^{91}$Nb, while two reactions produce it, $^{93}$Nb($\gamma$,n)$^{92}$Nb and $^{92}$Zr(p,n)$^{92}$Nb. A minor production channel (about 3\%) is $^{91}$Zr(p,$\gamma$). Because the two reactions destroying $^{92}$Zr -- $^{92}$Zr(p,n) and $^{92}$Zr(p,$\gamma$) -- both eventually lead to $^{92}$Nb production, their relative magnitude is not important, only their combination into a total rate. The production of $^{92}$Zr proceeds via ($\gamma$,n) sequences from the other Zr isotopes. The slowest reactions in these sequences are the ones removing a paired neutron and thus they dominate the timescale and the flow. Here, this is $^{94}$Zr($\gamma$,n) and, with slightly lower
importance, $^{96}$Zr($\gamma$,n), both leading to eventual production of $^{92}$Zr. Finally, $^{94}$Nb($\gamma$,n)$^{93}$Nb is important in the production of $^{93}$Nb from neutron-richer Nb isotopes.
The rates of $^{94}$Zr($\gamma$,n) and $^{94}$Nb($\gamma$,n)$^{93}$Nb are experimentally determined through their measured neutron capture cross sections \cite{kadonis} because despite of the elevated temperatures found in $\gamma$-process nucleosynthesis, the relevant stellar neutron capture rates are still dominated by the ground-state contributions \cite{rau12}. Thus, the experimental data constrains capture rates well in these cases \cite{rau14}. The photodisintegration rates are then also well constrained because they are computed from the capture by applying detailed balance \cite{rau00}. The $^{96}$Zr($\gamma$,n) rate comes from a theory estimate as given in \cite{bao}, including contributions from thermally populated excited states. For the other rates given above, and their reverse reactions, we used predictions by \cite{rau00} in our standard calculations.

The uncertainty in the $^{92}$Nb/$^{92}$Mo ratio also contains the uncertainty in the $^{92}$Mo production. Figure \ref{fig:flowMoNb} shows the time-integrated flows in the tracer that produces the main fraction of $^{92}$Mo. The flow pattern is less complex than in the case of $^{92}$Nb. The main contribution to this nuclide (about 50\%) is through ($\gamma$,n) sequences coming from the stable Mo isotopes with mass numbers $A>94$. These are mainly producing $^{94}$Mo, part of which is converted to $^{92}$Mo through the reaction sequence $^{94}$Mo($\gamma$,n)$^{93}$Mo($\gamma$,n)$^{92}$Mo. The slower reaction in this sequence, and thus determining the flow, is $^{94}$Mo($\gamma$,n), leaving an unpaired neutron in $^{93}$Mo. The second important path, contributing about 35\%, is the sequence $^{93}$Nb(p,n)$^{93}$Mo($\gamma$,n)$^{92}$Mo. Although the magnitude of the (p,n) reaction also scales with the proton density, the $^{93}$Mo($\gamma$,n) reaction is the faster one again in this sequence at our SNIa
conditions. Finally, the reaction $^{91}$Nb(p,$\gamma$) provides a small (15\%), additional contribution to $^{92}$Mo. There are only theoretical predictions available for the rates which are important, $^{94}$Mo($\gamma$,n), $^{93}$Nb(p,n), and (with lower impact) $^{91}$Nb(p,$\gamma$). The $^{92}$Mo production would scale according to the above weights when new rate determinations become available for these reactions.

The important, only theoretically determined rates affecting the production of $^{92}$Nd and $^{92}$Mo are summarized in Table \ref{tab:nbmoratio}. In order to check the uncertainty in our GCE value for the $^{92}$Nb/$^{92}$Mo ratio due to uncertainties in the reaction rates, we varied the most important rates found above. The resulting range, after GCE, is also shown in Table \ref{tab:nbmoratio}.

\section{Monte Carlo studies}


%
\begin{figure}
\includegraphics[width=0.5\columnwidth]{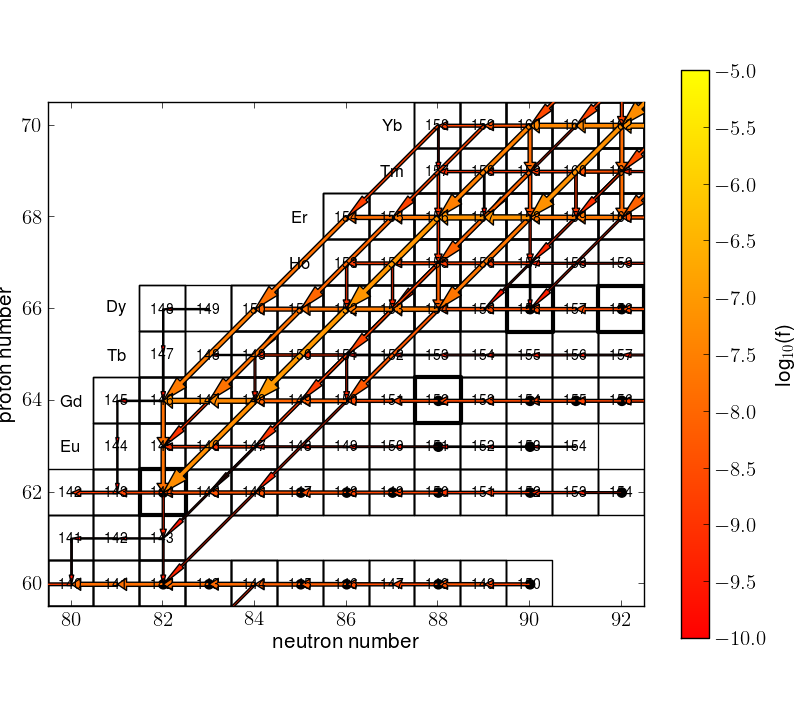}
\includegraphics[width=0.5\columnwidth]{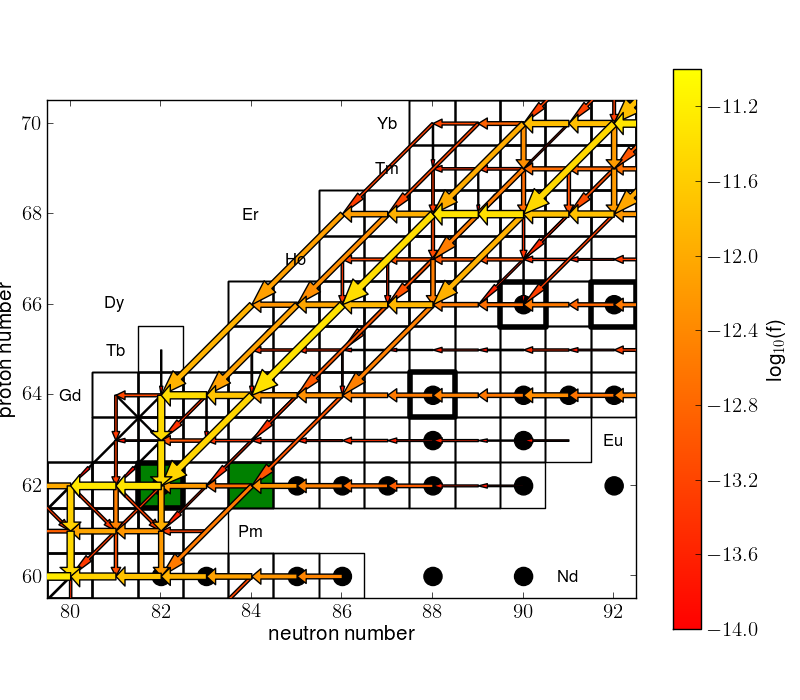}
\caption{\label{fig:sm}Reaction flow for producing $^{146}$Sm and $^{144}$Sm production; size and color of the arrows relate to the magnitude of the time-integrated flux on a logarithmic scale. The left panel shows the flows in the Sm-like tracer of the SNIa model, the right panel shows the flows in ccSN (summed over all $\gamma$-process zones).}

\end{figure}

\begin{figure}

\includegraphics[width=0.5\columnwidth]{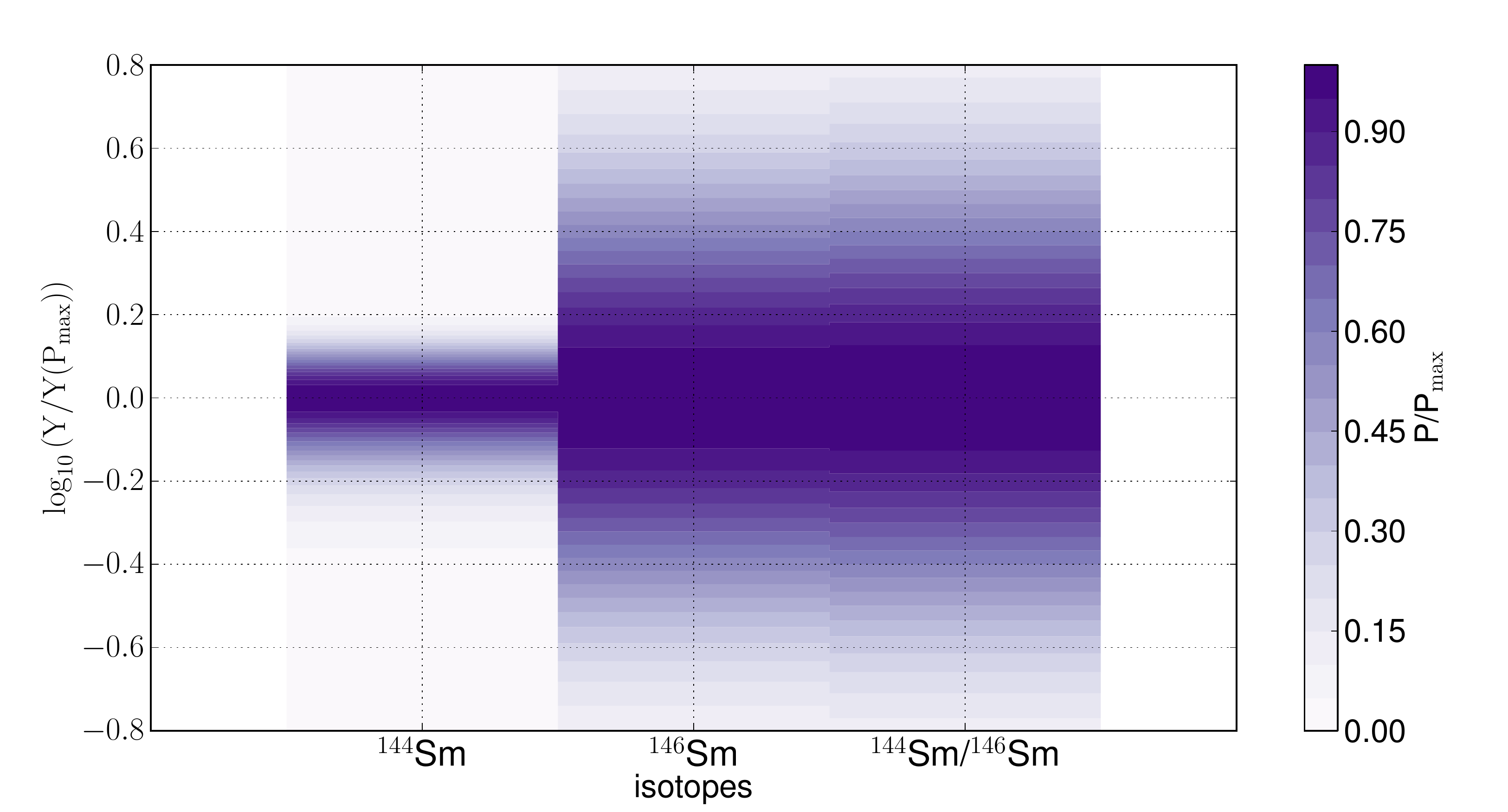}
\includegraphics[width=0.5\columnwidth]{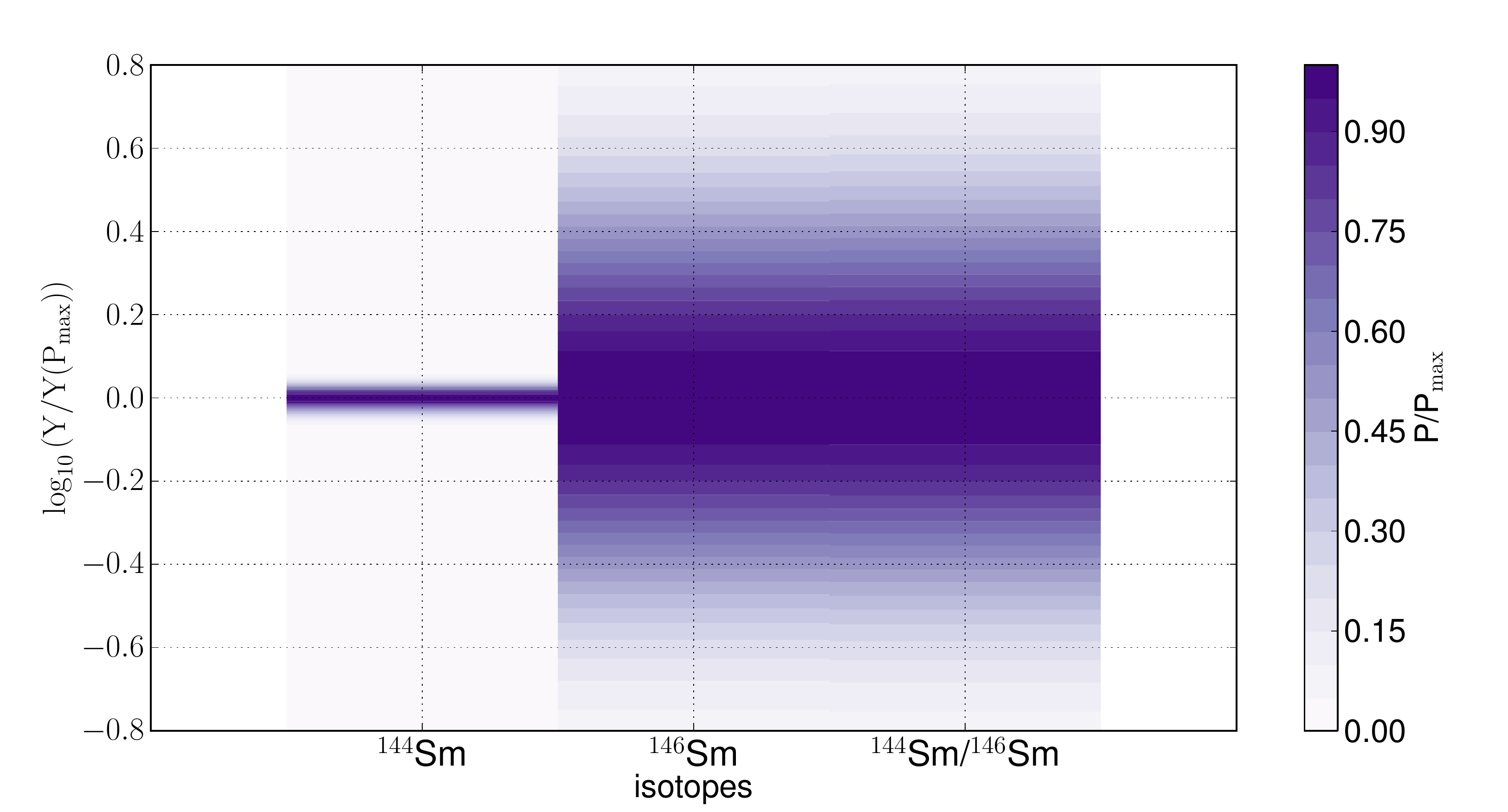}
\centerline{\includegraphics[width=0.6\columnwidth]{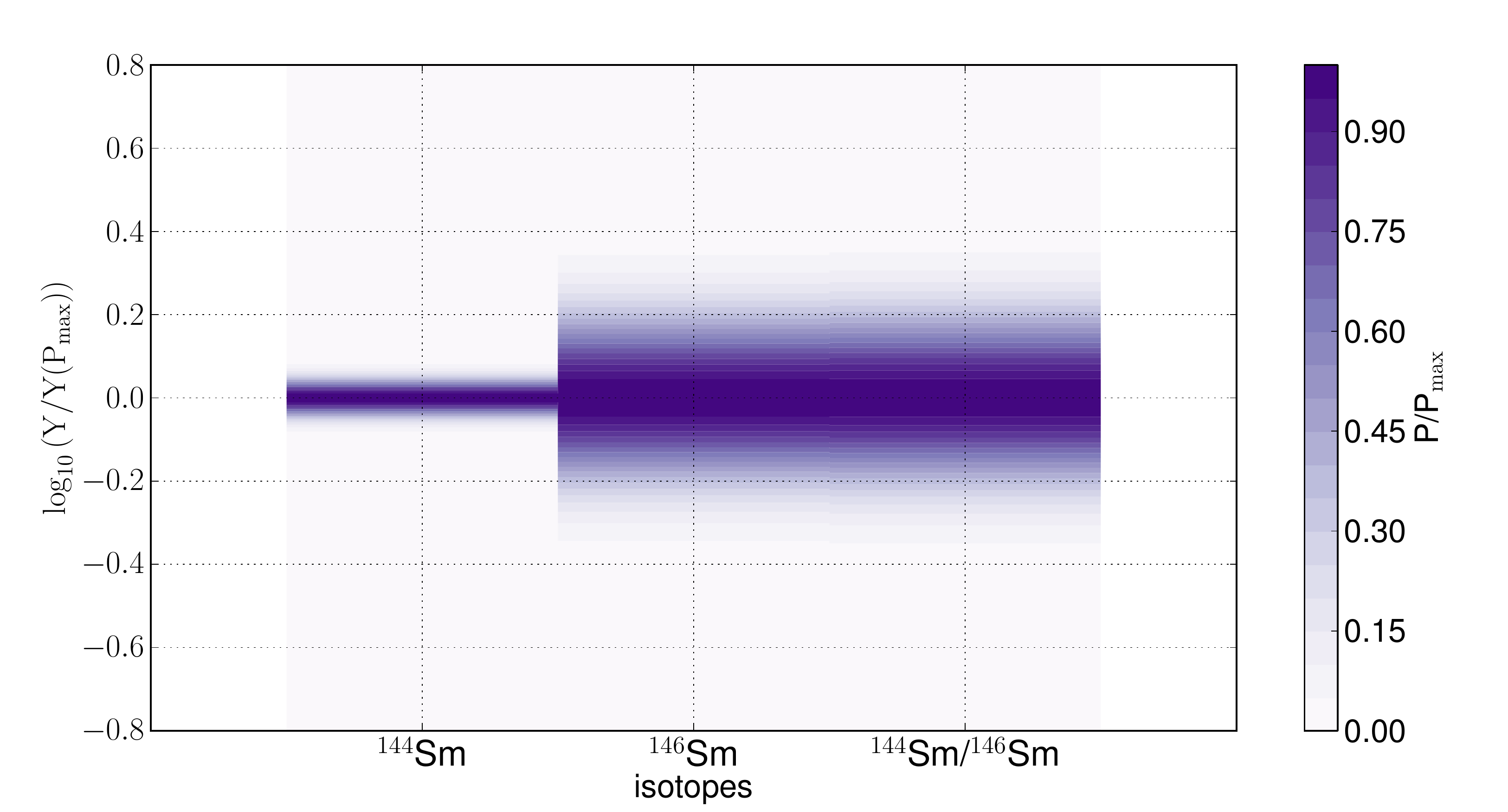}}
\caption{\label{fig:smdens}Each panel shows PDDs for $^{144}$Sm (left), $^{146}$Sm (center) and their ratio (right) obtained from Monte Carlo variations of all reactions in the SNIa tracer. Different uncertainty factors were used depending on the reaction and whether it is predicted or measured. The PDDs are for variation of all rates (top left panel), variation of ($\alpha$,$\gamma$) only (top right panel), variation of (n,$\gamma$) only (lower panel).}
\end{figure}

Additionally, we show the results of full Monte Carlo variations in the PizBuin framework \cite{pizbuin}, which combines a Monte Carlo driver with a fast, parallelized reaction network. Here we only show first exploratory calculations with reduced networks and simple variation factors but the setup will ultimately allow us to perform comprehensive, large-scale studies of nuclear uncertainties in abundance predictions, including 10000s of reactions with individual nuclear uncertainties.

Instead of identifying possibly important reactions in flow plots and varying their rates manually, as done in Sec.\ \ref{sec:paths}, in the Monte Carlo (MC) approach we varied all reactions or reactions of a specific type simultaneously to find their impact on final uncertainties in the calculated abundances of a given nuclide. When varying a rate, we assumed symmetric uncertainty factors of 1.3 and 2.0 for experimentally determined and theoretical neutron captures, respectively. Theory rates for reactions involving protons received an uncertainty factor of 3.0, whereas an asymmetric uncertainty was used for predicted rates with $\alpha$ particles. In the latter case, an uncertainty factor of 2.0 was assumed for the upper limit and a factor of 0.1 for the lower limit.

For the $^{92}$Nb and $^{92}$Mo isotopes, the MC variation of the full rate set and the reduced set shown in Table \ref{tab:nbmoratio} agrees excellently as seen in Fig.\ \ref{fig:dens}. This demonstrates the appropriateness of the selection of reactions from the flow plots as performed above but provides a better quantification of the uncertainties through probability density distributions (PDDs), for the individual nuclides as well as their abundance ratio.

A similar study was performed for the $^{146}$Sm and $^{144}$Sm isotopes, showing the advantage of the MC approach. The flow pattern shown in Fig.\ \ref{fig:sm} is more complicated and making a selection of few reactions impossible. The PDDs from the Monte Carlo variations shown in Fig.\ \ref{fig:smdens} demonstrate that the uncertainty of the $^{146}$Sm/$^{144}$Sm ratio is governed by those of ($\gamma$,$\alpha$) reactions on unstable, proton-rich nuclei. This is similar to the ccSN case.

\section{Summary}
In conclusion, the SNIa model provides a viable site to explain the radiogenic p-nuclides and their abundance ratios in the early solar system as derived from meteoritic abundances. The involved nuclear uncertainties are specifically large for the Sm isotopes, whereas the $^{92}$Nb/$^{92}$Mo ratio is predicted with smaller nuclear uncertainty. Monte Carlo approaches allow to quantify uncertainties also for complicated flow patterns with many contributing reactions.

\acknowledgments
This work is partially supported by the Swiss NSF, the European Research Council (EU-FP7-ERC-2012-St grant 306901; ERC Advanced Grant GA 321263-FISH), and the B2FH Association.

\end{document}